\documentclass[journal]{IEEEtran}

\usepackage{ifpdf}

\usepackage[space]{cite}

\usepackage[usenames]{color} 

\usepackage[utf8]{inputenc} 

\usepackage{mdwlist}

\usepackage{placeins}

\usepackage[draft]{pgf}

\usepackage{graphicx}
\DeclareGraphicsExtensions{.pdf,.jpeg,.png}

\usepackage{amssymb} 
\usepackage{amsmath}

\usepackage{booktabs}

\usepackage{algorithmic}
\usepackage{array}
\usepackage[caption=false,font=footnotesize]{subfig}
\usepackage{url}

\DeclareUrlCommand\email{\urlstyle{rm}}

\usepackage{hyperref} %

\begin{document}
\bstctlcite{IEEEexample:BSTcontrol}

\title{Forecasting intracranial hypertension using multi-scale waveform metrics}

\author{Matthias~H\"user,
        Adrian~K\"undig,
        Walter~Karlen,
        Valeria~De~Luca,
        and~Martin~Jaggi%
\thanks{M. H\"user is with the Biomedical Informatics Group, Department of Computer Science, ETH Z\"urich, Universitätstrasse 6, 8092 Z\"urich,
        Switzerland (correspondence e-mail: \texttt{matthias.hueser@inf.ethz.ch}). A. K\"undig was with the Department of Computer Science, 
        ETH Z\"urich, 8092 Z\"urich, Switzerland. W. Karlen is with the Mobile Health Systems Lab, Institute of Robotics and Intelligent Systems, 
        Department of Health Sciences and Technology, ETH Z\"urich, 8008 Z\"urich, Switzerland. V. De Luca was with the Department of Information 
        Technology and Electrical Engineering, ETH Z\"urich, 8092 Z\"urich, Switzerland. She is now with the Novartis Institutes for Biomedical Research, 4056 Basel, Switzerland.
        M. Jaggi is with the Machine Learning \& Optimization Lab, EPFL, 1015 Lausanne, Switzerland.}}

\markboth{}%
{Shell \MakeLowercase{\textit{H\"user et al.}}: Forecasting intracranial hypertension using multi-scale waveform metrics}

\maketitle

\begin{abstract}

Objective: Acute intracranial hypertension is an important risk factor of secondary brain damage after traumatic brain injury. Hypertensive episodes
are often diagnosed reactively, leading to late detection and lost time for intervention planning. A pro-active approach that predicts critical events several
hours ahead of time could assist in directing attention to patients at risk. Approach: We developed a prediction framework that forecasts onsets of acute intracranial hypertension in the next 8 hours. It jointly uses cerebral auto-regulation indices, spectral energies and morphological pulse metrics to describe the neurological 
state of the patient. One-minute base windows were compressed by computing signal metrics, and then stored in a multi-scale history, 
from which physiological features were derived. Main results: Our model predicted events up to 8 hours in advance with alarm recall rates
of 90\% at a precision of 30.3\% in the MIMIC-III waveform database, improving upon two baselines from the literature. We found that features derived 
from high-frequency waveforms substantially improved the prediction performance over simple statistical summaries of low-frequency time series, 
and each of the three feature classes contributed to the performance gain. The inclusion of long-term history up to 8 hours was 
especially important. Significance: Our results highlight the importance of information contained in 
high-frequency waveforms in the neurological intensive care unit. They could motivate future studies on pre-hypertensive patterns and the design of new alarm 
algorithms for critical events in the injured brain.

\end{abstract} 


\section{Introduction}

With at least 10 million cases annually leading to hospitalization worldwide, traumatic brain injury (TBI),
often causing intracranial hemorrhage, is a major public health issue \cite{langlois06_epidemiology_impact_traumatic}. 
After initial admission to the intensive care unit (ICU) and assessment of the primary brain injury, further neurological 
damage often occurs. This phenomenon is referred to as \emph{secondary brain injury}, and often leads to long-term brain damage
through e.g. cerebral ischemia (decrease of blood flow to the brain) \cite{bramlett04_pathophysiology_cerebral_ischemia}, 
cerebral hypoxia (decrease of substrate/oxygen flow to the brain) \cite{oddo11_brain_hypoxia_associated} and brain
herniation (swelling leading to compression of brain structures \cite{rehman08_rapid_progression_traumatic}).

\begin{figure}[h!]
\centering
\includegraphics[width=0.49\textwidth]{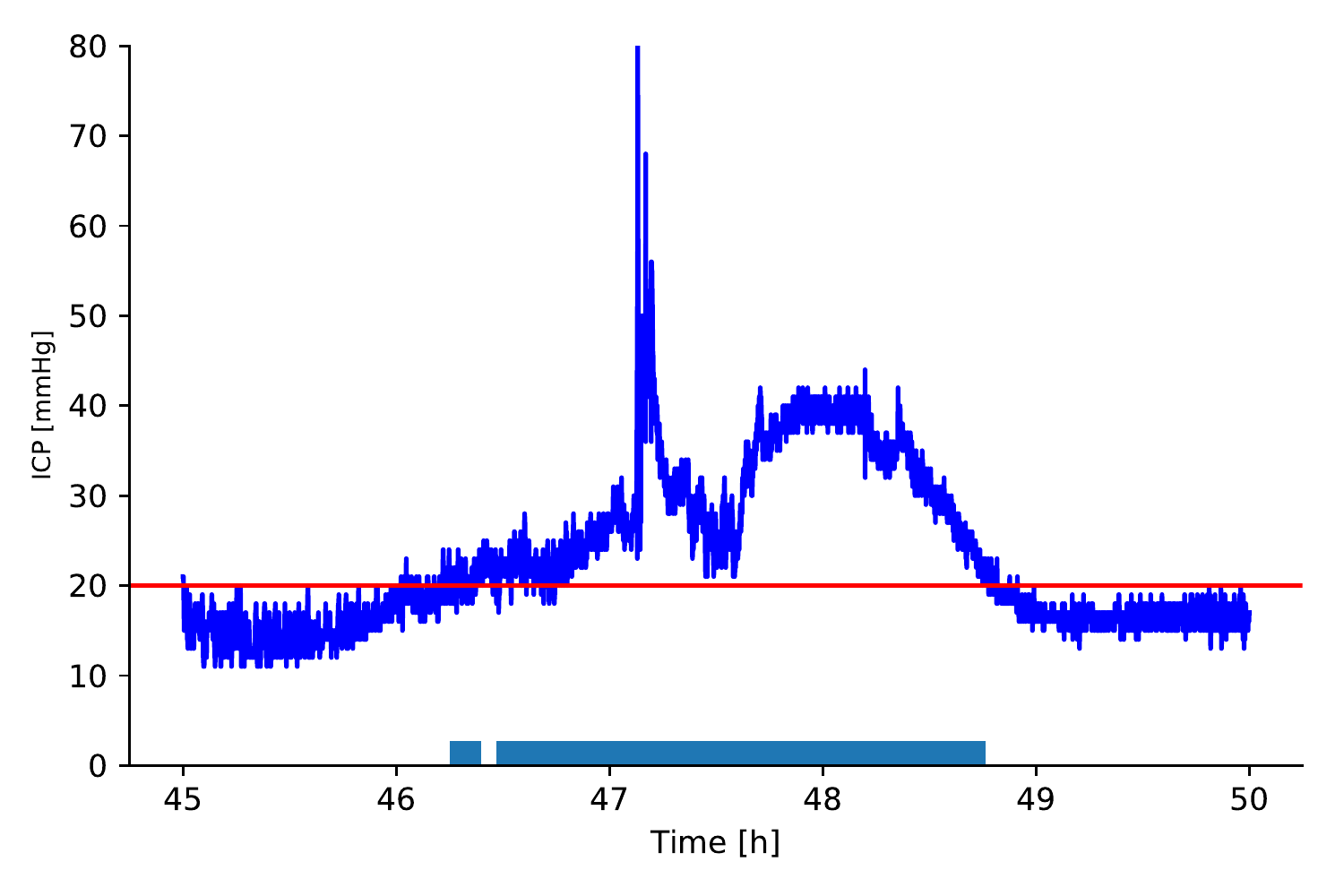}
\caption{ICP recording illustrating an acute intracranial hypertension event. The patient
         exhibits normal ICP values, before deteriorating into intracranial hypertension (ICH, boxes on rug plot).  
         At the end of the trace the patient returns to a normal ICP state. The horizontal red line denotes the threshold defining ICH.}
\label{fig:ichexample}
\end{figure}

Management of TBI patients in the neurological ICU focuses on mitigating and possibly reversing secondary
injuries \cite{werner07_pathophysiology_traumatic_brain}. A key variable in the management of secondary brain injury 
is \emph{intracranial pressure} (ICP) \cite{lavinio11_intracranial_pressure_why,carney17_guidelines_management_severe}.
Cerebral compliance maintains blood- and energy substrate flow by holding pressure constant against slight
volume changes of the cranial components \cite{rangelcastilla08_cerebral_pressure_autoregulation}. The ICP value of a healthy
 adult is maintained by this mechanism in the range 7-15 mmHg \cite{rangelcastilla08_management_intracranial_hypertension}. 
However, if compliance is reduced, rapid non-linear ICP elevations can occur \cite{mcnames01_precursors_rapid_elevations}. A sustained ICP 
elevation over 20 mmHg is defined as \emph{acute intracranial hypertension} (ICH) \cite{bardt98_monitoring_brain_tissue}. An illustrative 
example of an ICH event is shown in Fig.~\ref{fig:ichexample}.

A direct association of time spent in the ICH state with clinical outcome has been empirically shown: The area under the ICP curve in 
the first 48 hours of ICU treatment is an independent predictor of in-hospital mortality \cite{badri12_mortality_long_term}. Various
other studies have established an association of ICH and poor neurological outcome 
\cite{bonds15_predictive_value_hypothermia,karamanos14_intracranial_pressure_versus,majdan14_timing_duration_intracranial}.
Accordingly, it is a common treatment goal in neuro-critical care to avoid acute intracranial hypertension \cite{carney17_guidelines_management_severe}.
Invasive, intra-parenchymal ICP monitoring combined with interventions such as external ventricular drainage or surgery is
the gold standard to control and maintain ICP in the physiological range of 7-15 mmHg and ensure adequate cerebral compliance \cite{rangelcastilla08_management_intracranial_hypertension}. 
Advances in monitoring and signal processing technology have allowed to record high-frequency ICP traces and analyze them
in real-time \cite{bhatia07_neuromonitoring_intensive_care}. Yet there are several caveats that hinder the interpretation
of the ICP signal and its use as a decision-support tool: (a) raw data and time-varying trends are presented to the clinician, and no risk
estimates for ICH are available. This can lead to information overload and over-consumption of human attention for the ICU personell. For 
example, a study has found that clinicians are often not confident that effort spent on inspection of ICP traces is redeemed 
by improving outcome after TBI \cite{sahjpaul00_intracranial_pressure_monitoring}; (b) threshold-based track-and-trigger systems usually
have too high false alarm rates, which can desensitize staff to dangerous hypertension events \cite{chambrin99_multicentric_study_monitoring}; 
(c) alarms are only triggered after onset of acute intracranial hypertension, when long-term effects might be harder to prevent.

To address these problems, robust \emph{forecasting} of ICH onsets could augment the current treatment protocol 
which is reactive in nature. Previous works have shown that complex precursor patterns occur in auto-regulation 
indices and ICP/ABP waveform morphology prior to hypertensive events \cite{mcnames02_sensitive_precursors_acute,hu10_forecasting_icp_elevation}. 
Recently, simple prediction models explicitly targeting ICH forecasting 30 minutes up to 6 hours before the event were proposed, 
yielding promising results \cite{myers16_predicting_intracranial_pressure,guiza13_novel_methods_predict,guiza17_early_detection_increased}. 
However, it is not well understood or investigated which of these approaches is necessary or sufficient to achieve high prediction performance in
the context of an early warning system for ICH, and whether additional benefits could be derived from their combination.

In this extensive empirical study of ICH prediction we make the following contributions

\begin{itemize*}
\item An online ICH prediction framework which describes the neurological state using multi-scale metrics
      of the last 8 hours of recordings, comprising classical statistical features, cerebral auto-regulation indices, frequency band 
      energies and ICP/ABP pulse morphology computed on high-frequency waveforms. The resulting model is shown to outperform two baseline models from the 
      literature in a controlled comparison.
\item By including a wide range of relevant channels and physiological feature types, we conduct the first systematic study 
      of ICH prediction across signal channels and feature types, and thereby benchmark various pre-hypertensive
      patterns exploited or hypothesized in previous works.
\item We demonstrate clear performance benefits when including morphological and spectral energy features derived from high-frequency waveforms 
      compared to focusing on only statistical metrics on low-frequency time series which have been often used in major recent works.
\item Using the state-of-the-art feature attribution technique SHAP (SHapley Additive exPlanation) we study the importance 
      and generate rankings of different features that explain positive intracranial hypertension alarms, representing
      the first application of this technique to an extensive set of pre-hypertensive patterns.
\end{itemize*}

Preliminary and partial versions of this work have been reported in clinical abstracts \cite{de2016temporal,huser2015forecasting}.
\section{Related work}

The association of information contained in high-frequency physiological waveforms/time series and elevated ICP
has been studied in various works. For example, Hornero et al. \cite{hornero05_interpretation_approximate_entropy} have found that decreased
ICP signal complexity and irregularity is associated with intracranial hypertension. Fan et al. \cite{fan10_approach_determining_intracranial}
identified an association between ICP variability and decreased pressure auto-regulation. Recently, it was established that
characteristic patterns in various physiological channels are correlated with ICP and could thus be used to predict ICH \cite{naraei17_towards_learning_intracranial}.
Several auto-regulation indices defined on physiological channels were reported, such as by Zeiler et al. ~\cite{zeiler18_description_new_continuous}, which
studied the moving correlation coefficient between ICP, ABP and CPP channels, and others \cite{aries12_continuous_monitoring_cerebrovascular,radolovich11_pulsatile_intracranial_pressure}. 
The relationship between auto-regulation indices and successive ICH events has been studied by Kim et al. \cite{kim16_trending_autoregulatory_indices}. In 
general, it has long been suspected that the information contained in the pulsatile ICP signal is very rich beyond simple statistical
summaries \cite{balestreri04_intracranial_hypertension,kirkness00_intracranial_pressure_waveform}.

Besides auto-regulation indices, previous works have attempted to use morphological descriptors of the intracranial pressure pulse to 
predict ICH onset up to 20 minutes in advance \cite{scalzo12_intracranial_hypertension_prediction,hu10_forecasting_icp_elevation,hamilton09_forecasting_intracranial_pressure}. 
More generally, morphological analysis of ICP pulses \cite{eide06_new_method_processing} has emerged as a successful approach
and was used to e.g. reduce false alarm rates of ICP alarms \cite{scalzo13_reducing_intracranial_pressure} and
track pulse metrics in real-time \cite{scalzo12_intracranlal_pressure_signal}. Hu et al. \cite{hu09_morphological_clustering} applied cluster analysis
to individual ICP pulses. Other types of features that have been proposed to summarize physiological time series
include bag-of-words of physiological motifs applied to ECG/EEG time series \cite{wang13_bag_words_representation} and
entropy measures \cite{xu13_improved_wavelet_entropy,lu12_complexity_intracranial_pressure}. The recently proposed ICP trajectories framework 
\cite{jha18_intracranial_pressure_trajectories} uses longitudinal ICP time series to discover clinical phenotypes. Different approaches have also been proposed, based on assessing risk only from static clinical data \cite{pace18_clinical_prediction_model} or biomarkers \cite{stein12_use_serum_biomarkers,adamides09_brain_tissue_lactate,hergenroeder08_identification_serum_biomarkers}, instead of using historical time series.

To obviate the need for explicit feature engineering on historical time series, deep learning architectures have been proposed, which detect intracranial hypertension from the 
raw pulse waveform \cite{quachtran16_detection_intracranial_hypertension}. Simpler dimensionality reduction approaches, such as principal component analysis, have also been used to find non-correlated features \cite{naraei17_pca_based_feature} that describe ICH.

Major recent works explicitly addressing the ICH forecasting problem include the approach proposed by G\"uiza et al. \cite{guiza13_novel_methods_predict}, which obtained 
an AUROC of 0.87 for prediction of ICH in the next 30 minutes. Their analysis showed that the most predictive channel is
ICP and that the most recent measurements are the most relevant features. Subsequently, their model was externally validated, resulting in similar performance \cite{guiza17_early_detection_increased}. 
Myers et al. \cite{myers16_predicting_intracranial_pressure} proposed a model that is able to predict ICH
up to 6 hours in advance, a prediction horizon comparable to our method. It uses simple features such as the last measured ICP value or the time to 
the last ICH crisis. Besides tackling the classification task directly, other models have been suggested that predict the future 
ICP mean value, for example by using nearest-neighbor regression~\cite{bonds15_predicting_secondary_insults}, neural networks \cite{zhang11_artificial_neural_network,shieh04_intracranial_pressure_model} or ARIMA models \cite{zhang12_online_icp_forecast}.
\section{Methods}
\label{sec:meth}

\subsection{Physiological database}

In all experiments, we have used the multi-parameter intelligent monitoring in intensive care III
waveform database (MIMIC-III WFDB) \cite{saeed11_multiparameter_intelligent_monitoring}, Version 1.0. The entire 
dataset consists of 67,830 records
extracted from patient stays at the Beth Israel Deaconess Medical Center, Boston, MA, United States. The MIMIC-III WFDB was 
chosen for this study because it 
contains simultaneous measurements of high-frequency waveforms (125 Hz) and derived time series (1 Hz) for a range of physiological channels that are
relevant to the prediction problem. Among all available channels, we selected \texttt{ICP} (mean intracranial pressure) , 
\texttt{CPP} (cerebral perfusion pressure), \texttt{ABPm/d/s} (mean/diastolic/systolic arterial blood pressures), and
\texttt{HR} (heart rate) time series, and \texttt{wICP} (intracranial pressure), \texttt{wABP} (invasive arterial 
blood pressure), \texttt{wPLETH} (raw output of fingertip plethysmograph), \texttt{wRESP} (respiration waveform) 
and \texttt{ECG} waveforms (Fig. \ref{fig:methods-overview}a). This broad range allows us to compare the relevance of
different channels for predictive modeling, while ensuring that we can extract a cohort of at least 100 ICU recordings with regular sampling. 
Waveforms were acquired using the bedside IntelliVue Patient Monitoring system, Philips Healthcare, The Netherlands.

\subsection{Cohort selection}

\begin{figure}[h!]
\centering
\includegraphics[width=0.49\textwidth]{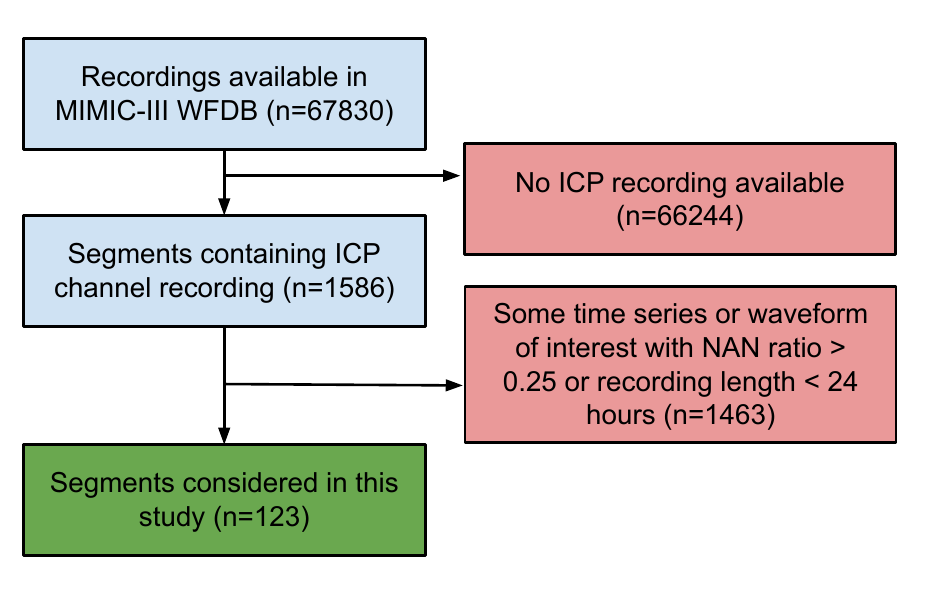}
\caption{Overview of selection criteria that were used to define the recordings of interest for this study}
\label{fig:pat-cohort-inclusion}
\end{figure}

Only a small fraction of available records in the MIMIC-III WFDB contain ICP data. In a first step, we discarded all segments that have
no available \texttt{ICP} time series, which left 1586 relevant segments. We further require a minimum
recording length of 24 hours, and a missing value ratio of at most 25\% for each
considered waveform or time series channel. We applied these criteria to ensure that the relevance of different channels as features could
be meaningfully compared, and individual channels would not be negatively affected by long stretches of missing data. After applying these criteria, 
123 segments remained in the cohort. This set of recording segments was used in all reported experiments. A diagram summarizing patient exclusions and cohort definition
is shown in Fig. \ref{fig:pat-cohort-inclusion}. Matching to MIMIC-III clinical database records was only possible for 66 of 
the 123 segments, contributed by 50 unique patients, for which clinical context 
is provided in Table \ref{tab:cohort-demographics}. In terms of admission diagnosis, the cohort is 
homogeneous, with most patients exhibiting intracranial hemorrhage. In this work, we did not include
clinical covariates into the processing pipeline to avoid unequal treatment of segments or a significant reduction of available segments. 
Overall, our data-set contains 10,547 hours of data (Fig. \ref{fig:methods-overview}a). Each segment has a mean recording 
length of 85.8 hours (std: 54.8 hours). The mean ICP value in the cohort is 9.1 mmHg (std: 6.8 mmHg). 

\begin{table}[h!]
\caption{Patient demographic information for ICU stays ($n_s$=50) matched with $n$=66 recording segments, $n$=57
         recording segments could not be matched in the MIMIC-III CDB}
\vspace{2mm}
\centering
\resizebox{0.49\textwidth}{!}{%
\begin{tabular}{ll}
\toprule
\textbf{Age median [IQR]} & 62.5 [57.0-72.5] \\ 
\textbf{Sex (\% male)} & 38.0 \\ 
\textbf{Hospital mortality rate (\%)} & 22.0 \\ 
\textbf{ICU LOS days median [IQR]} & 13.9 [7.4-21.7] \\ 
\textbf{Admission type} & Emergency ($n_s=48$) \\ 
               & Elective ($n_s=2$) \\
\textbf{ICU care service}   & Neurological surgical ($n_s=42$) \\ 
               & Neurological medical ($n_s=3$) \\
               & Unspecified medical ($n_s=3$) \\
               & Cardiac medical ($n_s=1$) \\
               & Obstetric ($n_s=1$) \\
\textbf{Diagnosis} & Subarachnoid hemorrhage ($n_s=21$) \\ 
                   & Intracranial hemorrhage ($n_s=19$) \\ 
                   & Interparenchymal hemorrhage ($n_s=2$) \\ 
                   & Brain tumor ($n_s=2$) \\ 
                   & Headache ($n_s=1$) \\
                   & Bleed($n_s=1$) \\
                   & Other (hematology) ($n_s=1)$ \\
                   & Other (respiratory) ($n_s=1$) \\
                   & Other (hepatology) $(n_s=1)$ \\
                   & Other (obstetric) ($n_s=1$)\\
\textbf{GCS median} [IQR] & 3 [3-4] \\
\bottomrule
\end{tabular}}
\label{tab:cohort-demographics}
\end{table}

\subsection{Acute intracranial hypertension alarms}

\begin{figure*}[h!]
\centering
\includegraphics[width=0.98\textwidth]{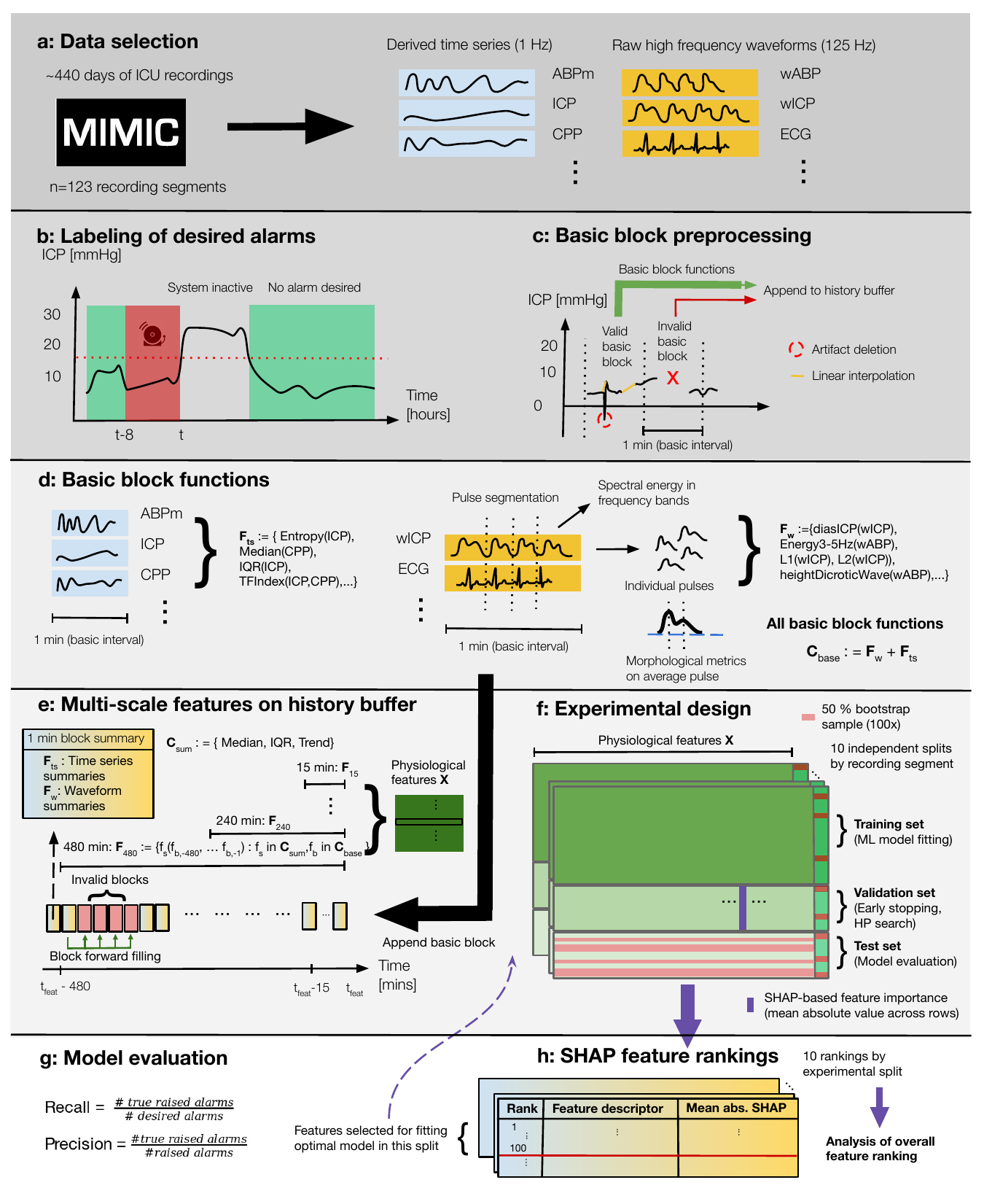} 
\caption{Overview of physiological feature extraction, alarm labeling and experimental design of the performed study}
\label{fig:methods-overview}
\end{figure*}

We define an ICH event as 5 successive 1-minute blocks with median ICP greater than 20 mmHg, which helps to avoid 
spurious labelings, following the recent work by Ziai et al. \cite{ziai19_intracranial_hypertension_cerebral}, the
earlier work by Hu et al. \cite{hu10_forecasting_icp_elevation}, and the definition of sustained ICH
from \cite{marik99_management_increased_intracranial}. According to this definition, patients were in an acute ICH event
state for 2.5\% of their cumulative segment lengths. When analyzed by segment, the ICH state was active
for a mean of 2.0\% (std: 7.3\%) of each segment's
duration. A time point on the 1-minute grid was labeled as positive if there is \emph{any acute intracranial hypertension} event in the next 8 hours
and the patient is not already hypertensive (Fig. \ref{fig:methods-overview}b). This strategy implements an early warning system deployed in phases 
where the patient has normal ICP values.  Our design choice was to train one overall model for predicting events in the next 8 hours, without targeting
any specific prediction horizon. Positive 
labels correspond to time points at which an alarm should be produced by our prediction model (Fig. \ref{fig:methods-overview}b). Recall is defined 
as the fraction of those points, at which an alarm is indeed produced. Precision denotes the fraction of produced alarms which are in the 8 hours
 prior to some ICH event (Fig. \ref{fig:methods-overview}g). Both metrics are maximized if continuous sequences of alarms are 
 produced \emph{exactly} in the 8 hours before events, one for each grid point.
However, in clinical implementation this strict condition could be relaxed by applying post-processing such as moving window functions over
the sequence of thresholded prediction scores. We consider such processing to be out-of-scope here, but we suspect it can improve practical alarm 
system performance significantly, both in terms of recall and false alarm rate. There were a total of 555,644 labeled 1-minute time points, 
of which 117,954 were positive, and 437,690 negative.

\subsection{Physiological feature extraction framework}

\subsubsection*{Basic block functions}

During the feature generation process, so-called \emph{basic block functions} are computed online on non-overlapping 
windows containing 1 minute of high-frequency waveforms/time series, corresponding to 60 samples @1 Hz or 7500 samples @125 Hz (Fig. \ref{fig:methods-overview}d). The
choice of 1 minute as
a basic interval makes computation of complex morphological functions tractable, increases robustness to signal artifacts and sensor detachments, and allows
to produce updated predictions every minute. Before computing basic
block functions, a window is pre-processed by removing physiologically implausible values. If at least half of the samples are valid, we
reconstruct the remaining samples by linear interpolation. Otherwise, invalid basic blocks marked by a symbolic value are emitted (Fig. \ref{fig:methods-overview}c).
Basic block functions are then computed on valid blocks. As basic block functions we have considered statistical/complexity summaries (median, interquartile range(IQR), line length, Shannon entropy),
spectral band energies of waveforms, morphological pulse summaries of the \texttt{wABP} and the \texttt{wICP} waveforms, as well as cerebral 
auto-regulation indices. Morphological pulse metrics are computed by an algorithm consisting of several
steps (Fig. \ref{fig:methods-overview}d). First, individual pulses on \texttt{wABP}/\texttt{wICP} are segmented, using variants of known algorithms 
\cite{pan85_realtime_qrs_detection,hu08_algorithm_extracting_intracranial,zong03_open_source_algorithm},
with the help of the \texttt{ECG} waveform as a reference to identify pulse onsets. Valid pulses in the window are then temporally scaled to make their lengths comparable, overlaid and 
averaged point-wise, yielding an averaged pulse. Morphological pulse metrics, modeled on those described by Hu et al. \cite{hu10_forecasting_icp_elevation} and 
Almeida et al. \cite{almeida13_machine_learning_techniques}, are then computed on the averaged pulse. A complete overview of the basic block functions is
provided in Table \ref{table:feat-function-overview}.

\begin{table}[h!]
\caption{Overview of basic block functions computed on 1-minute windows}
\centering
\resizebox{0.49\textwidth}{!}{%
\begin{tabular}{l}
\midrule
\textbf{Statistical/complexity summaries} (\texttt{ICP}, \texttt{CPP}, \texttt{ABPm/d/s}, \texttt{HR})\\
\midrule
Median, Interquartile range, Line length \cite{esteller01_line_length_efficient}, Shannon entropy \\
\midrule
\textbf{Spectral band energy metrics} (\texttt{wICP}, \texttt{wABP}, \texttt{wPLETH}, \texttt{wRESP}, \texttt{ECG}) \\
\midrule
Energy in frequency bands [0,1],[1,2],[2,3],[3,6],[6,9],[9,12],[12,15] Hz
 \\
\midrule
\textbf{Autoregulation indices on time series (1 Hz sample rate)} \\
\midrule
AmpIndex(\texttt{ICP},\texttt{ABPm}), AmpIndex(\texttt{ICP},\texttt{CPP}), AmpIndex(\texttt{CPP},\texttt{ABPm})  \cite{radolovich11_pulsatile_intracranial_pressure} \\
PaxIndex(\texttt{ICP},\texttt{CPP},\texttt{ABPm}) \cite{radolovich11_pulsatile_intracranial_pressure} \\
PrxIndex(\texttt{ICP},\texttt{CPP},\texttt{ABPm}) \cite{steiner02_continuous_monitoring_cerebrovascular,czosnyka97_continuous_assessment_cerebral} \\
RapIndex(\texttt{ICP},\texttt{CPP}) \cite{avezaat79_cerebrospinal_fluid_pulse,kim09_index_cerebrospinal_compensatory} \\
SlowWaveIndex(\texttt{ICP}) \cite{lemaire02_slow_pressure_waves} \\
TFIndex(\texttt{ICP},\texttt{ABPm}), TFIndex(\texttt{ICP},\texttt{CPP}), TFIndex(\texttt{CPP},\texttt{ABPm}) \cite{zhang98_transfer_function_analysis} \\
\midrule
\textbf{Autoregulation indices on waveforms (125 Hz sample rate)} \\
\midrule
AmpIndex(\texttt{wICP},\texttt{wABP}) \cite{radolovich11_pulsatile_intracranial_pressure} \\
SlowWaveIndex(\texttt{wICP})  \cite{lemaire02_slow_pressure_waves} \\
TFIndex(\texttt{wICP},\texttt{wABP})  \cite{zhang98_transfer_function_analysis} \\
IaacIndex(\texttt{wICP},\texttt{wABP}) \cite{eide06_new_method_processing} \\
\midrule
\textbf{Morphological pulse metrics on waveforms} \\
\midrule
\textbf{wABP pulse descriptor} (17 metrics) \cite{almeida13_machine_learning_techniques}: \\
                                  A, UpstrokeTime, TimeAt$\Pi$, TimeAtDw, DownstrokeTime, \\
                                  SysDiasTimeDifference, HeightSysPeak, \\
                                  HeightInflPoint, HeightDicroticWave, \\
                                  R1, R2, R3, R4, R5, R6, Aix \\ 
\textbf{wICP pulse descriptor} (20 metrics) \cite{hu10_forecasting_icp_elevation}: \\ 
                                  Mean, Dias, DP1, DP2, DP3, DP12, DP13, DP23, \\
                                  L1, L2, L3, L12, L13, L23, Curv1, Curv2, \\
                                  Curv3, Slope, DecayTimeConst, AverageLatency
\end{tabular}}
\label{table:feat-function-overview}
\end{table}

\subsubsection*{Multi-scale history}

Computed basic block features are appended to a \emph{history buffer} using an online algorithm, with one batch
of features appended per minute (Fig. \ref{fig:methods-overview}e). If a block is invalid or some features cannot be computed, for example due to missing
signals, they are forward filled from the last valid feature in the history. If there is no valid feature in the recent past, 
the feature value is set to the median of that feature value in the accumulated history. After the history buffer 
is updated, a new sample of machine learning features is emitted by summarizing the current state of the history buffer (Fig. \ref{fig:methods-overview}e). As summary
functions, we use the median (location estimate of a basic feature over the history), IQR (variability of a basic feature over the history) and
the slope of regression line fit (trend of a basic feature over the history). These summary functions are applied 
separately over the last 15, 30, 60, 120, 240, 360 and 480 minutes to capture pre-hypertensive patterns at various scales
of the feature buffer history. After the full feature matrix is built, we standardize all feature
columns to have zero mean and unit standard deviation, using statistics from the training data-set. Missing values are 
replaced by zero, which corresponds to global mean imputation. For machine learning models that can deal with missing data natively, like
decision trees or tree ensembles, missing data imputation/normalization was not performed. The online signal processing and feature generation
algorithms were implemented using the numerical packages \texttt{SciPy} and \texttt{NumPy} in Python 3.6. 

\subsection{Feature interpretation using SHAP values}

To gain insights into the precursor patterns of intracranial hypertension we have used SHAP value analysis 
\cite{lundberg17_unified_approach_interpreting} to uncover the most important features that explain ICH 
predictions. SHAP (SHapley Additive exPlanation values) is a local feature attribution method, which attributes risk
scores of future intracranial hypertension to individual signal patterns, encoded in the
physiological features. The SHAP value of a feature $x_i=k$ is the expected change of the risk score, when this
feature is added to update the risk estimate, integrating over all possible subsets of other variables which are already
used in the risk estimate, prior to adding the new variable. The SHAP value of a prediction at $\mathbf{x}$ for feature $i$ 
is defined as $s_{i,\mathbf{x}} := E_S[ E[f(\mathbf{x})| \mathbf{x}_{S \cup \{i\}}] - E[f(\mathbf{x}) | \mathbf{x}_S]]$.
Here $f$ denotes the risk score,  $E[f(x)|\textbf{x}_S]$ the conditional expectation of the risk 
score if the values of features in $S$ are fixed to their observed values, and $E_S[\cdot]$ the expectation over the choice of fixed features $S$.
The used \texttt{TreeSHAP} algorithm \cite{lundberg18_consistent_individualized_feature} is an implementation of SHAP values for 
tree ensembles, which can deal with missing values, and hence simplifies the computations of $s_{i,\mathbf{x}}$. 
We summarized SHAP values of predictions on the validation set by defining the global importance of a feature as
$g_i := n^{-1}_\text{val} \sum_{\mathbf{x} \in \text{val}} |s_{i,\mathbf{x}}|$, as the mean magnitude of risk
score change that a particular feature causes when introduced into the model (Fig. \ref{fig:methods-overview}f). Hereby, $n_\text{val}$ is the number
of samples in the validation set. All features were ranked by $\{g_i\}$ for each
split, defining the \emph{top features} of the split. Ranks were averaged across splits to increase robustness
of the reported feature rankings (Fig. \ref{fig:methods-overview}g). SHAP values were also used as a \emph{feature selection} method internal to each split, 
by discarding all but the top 100 features on the validation set before creating derived models (Fig. \ref{fig:methods-overview}g). 
In this way, overfitting to non-informative features, which are numerous due to the broad range of feature combinations, is reduced. 
The test set was not used for feature selection or feature ranking purposes.

\subsection{Machine learning models}

As machine learning models we have considered \texttt{LogReg}, a L2-regularized logistic regression model optimized using stochastic gradient 
descent \cite{bottou10_large_scale_machine}; \texttt{Tree}, a single decision tree; \texttt{GradBoost}, a 
gradient-boosted ensemble of decision trees \cite{guyon17_lightgbm_highly_efficient}; and \texttt{MLP}, a multi-layer perceptron with a 
sigmoid activation function. Implementation details and hyper-parameter search grids for all machine
learning models are listed in the supplementary material.

\subsection{Experimental design}

Prediction models were evaluated using precision @75 and @90\% recall, which reflects our prior belief
that an alarm system for ICH should have high sensitivity, whereas false alarms are more tolerable and can be reduced
with post-processing defined on top of the sequence of prediction scores (Fig. \ref{fig:methods-overview}g). All experimental
result tables report these 2 metrics as well as areas under the PR curve. 95\% standard-error-based confidence intervals 
of performance metrics, which are used in all figures/tables, were constructed by drawing 10 randomized train/validation/test splits
(proportion 40:20:40\%) with respect to complete recording segments.
Splits were stratified, such that the positive label prevalence of training, validation, test sets in each split is within 0.015 
of the overall prevalence in the cohort. This minimizes nuisance effects for performance metrics sensitive to label prevalence (e.g. precision). 
The experiments performed per split are completely independent. The training set was used for model fitting, while the 
validation set was used for choosing the optimal set of hyperparameters, and computing mean absolute 
SHAP values that define the reported feature rankings (Fig. \ref{fig:methods-overview}f). Each split is associated with a distinct feature ranking,
which we integrate over in the feature importance analysis (Fig. \ref{fig:methods-overview}h). The test set was used to compute all reported performance
metrics; hyperparameters and optimal features were \emph{not} selected on this set to avoid overfitting. 
To account for test set variability, besides training process variability, we drew 100 bootstrap samples (size 50\% of test-set samples) 
with replacement from the test set, yielding 1000 overall replicates. Models with (indistinguishable based on overlapping 95\% confidence intervals) best performance are listed in bold-face.
\section{Results}

\subsection*{Low-frequency time series channels}

As a sanity check, we trained several models that do not use any features derived from high-frequency waveforms.
The results, shown in the first part of Table \ref{table:benefit-ts-wave}, indicate that \texttt{ICP} is the single most 
valuable time series across all desired recall levels. The addition of \texttt{ABP}/\texttt{CPP} context information leads to consistent performance increases 
for most evaluation metrics.

\begin{table}[h!]
\caption{Prediction performance of models by inclusion of physiological time series/waveform channels
         in the feature generation process}
\centering
\resizebox{0.49\textwidth}{!}{%
\begin{tabular}{llll}
\textbf{Channels} & \textbf{Prec@75Rec} & \textbf{Prec@90Rec} & \textbf{AUPRC} \\
\toprule
\textbf{ICP} & 0.311 $\pm$ 0.004 & N/A & 0.462 $\pm$ 0.007 \\ 
\textbf{ABP} & 0.226 $\pm$ 0.001 & 0.226 $\pm$ 0.000 & 0.238 $\pm$ 0.001 \\ 
\textbf{CPP} & 0.230 $\pm$ 0.001 & 0.223 $\pm$ 0.001 & 0.243 $\pm$ 0.003 \\ 
\textbf{ICP/ABP/CPP (1 Hz)} & 0.332 $\pm$ 0.003 & 0.267 $\pm$ 0.001 & 0.443 $\pm$ 0.005 \\ 
\midrule
\textbf{+wICP} & 0.371 $\pm$ 0.001 & \textbf{0.303} $\pm$ 0.001 & \textbf{0.512} $\pm$ 0.003 \\ 
\textbf{+wICP/ABP} & \textbf{0.377} $\pm$ 0.001 & \textbf{0.303} $\pm$ 0.001 & \textbf{0.517} $\pm$ 0.003 \\ 
\textbf{+wALL} & \textbf{0.379} $\pm$ 0.002 & \textbf{0.302} $\pm$ 0.001 & 0.510 $\pm$ 0.003 \\
\textbf{only wICP} & 0.358 $\pm$ 0.001 & 0.299 $\pm$ 0.001 & \textbf{0.516} $\pm$ 0.003 \\ 
\textbf{only wICP/wABP} & 0.366 $\pm$ 0.001 & 0.298 $\pm$ 0.002 & \textbf{0.516} $\pm$ 0.003 
\end{tabular}}
\label{table:benefit-ts-wave}
\end{table}

\subsection*{Importance of high-frequency waveform metrics}

Taking the most performant time series model (from the first part of Table \ref{table:benefit-ts-wave}), we tested whether adding features derived from 
125 Hz waveforms has a positive effect on the prediction performance (second part of Table \ref{table:benefit-ts-wave}).
Our results indicate that adding \texttt{wICP} yields a marked performance increase, and the joint use with \texttt{wABP}
strengthens this effect slightly. Using only waveform channels shows consistently higher performance than just using 
time series.

\subsection*{Morphological, spectral energy metrics and cerebral auto-regulation indices}

Morphological pulse metrics, cerebral auto-regulation indices and band energy have each
been shown to exhibit characteristic changes before hypertensive events in prior work. 
We tested whether such changes can translate into performance benefits when the corresponding 
features are added to a simple model. Our results are summarized in the first part of Table~\ref{table:base-summary-functions}. Incremental additions of 
feature categories (ordered roughly by computational cost and algorithmic complexity) lead to consistent performance increases across all
desired recalls.

\begin{table}[h!]
\caption{Prediction performance by models based on different basic block functions (first part), and multi-scale history summary functions (second part)}
\centering
\resizebox{0.49\textwidth}{!}{%
\begin{tabular}{llll}
\textbf{Feature types} & \textbf{Prec@75Rec} & \textbf{Prec@90Rec} & \textbf{AUPRC} \\
\midrule
\textbf{Stat/Complexity} & 0.328 $\pm$ 0.003 & 0.270 $\pm$ 0.002 & 0.464 $\pm$ 0.005 \\ 
\textbf{+BandEnergy} & 0.356 $\pm$ 0.002 & 0.286 $\pm$ 0.001 & 0.502 $\pm$ 0.003 \\ 
\textbf{+AutoRegIndices} & 0.368 $\pm$ 0.001 & 0.289 $\pm$ 0.001 & 0.503 $\pm$ 0.003 \\ 
\textbf{+PulseMorphology} & \textbf{0.377} $\pm$ 0.001 & \textbf{0.303} $\pm$ 0.001 & \textbf{0.517} $\pm$ 0.003 \\ 
\midrule
\textbf{Location} & 0.373 $\pm$ 0.002 & \textbf{0.300} $\pm$ 0.002 & 0.508 $\pm$ 0.004 \\ 
\textbf{Loc+Trend+Variation} & \textbf{0.377} $\pm$ 0.001 & \textbf{0.303} $\pm$ 0.001 & \textbf{0.517} $\pm$ 0.003  
\end{tabular}}
\label{table:base-summary-functions}
\end{table}

\subsection*{Multi-scale history summary modes}

It has been reported in the literature that variability or trends of individual 
metrics are predictive of ICH events. Using different history buffer summary functions, we tested whether such features are indeed valuable vs. location estimates. 
Our results, listed in the second part of Table \ref{table:base-summary-functions}, suggest that adding trend/variability functions to the 
multi-scale history provides benefits.

\subsection*{How much history do we need to store?}

\begin{figure}[h!]
\centering
\includegraphics[width=0.49\textwidth]{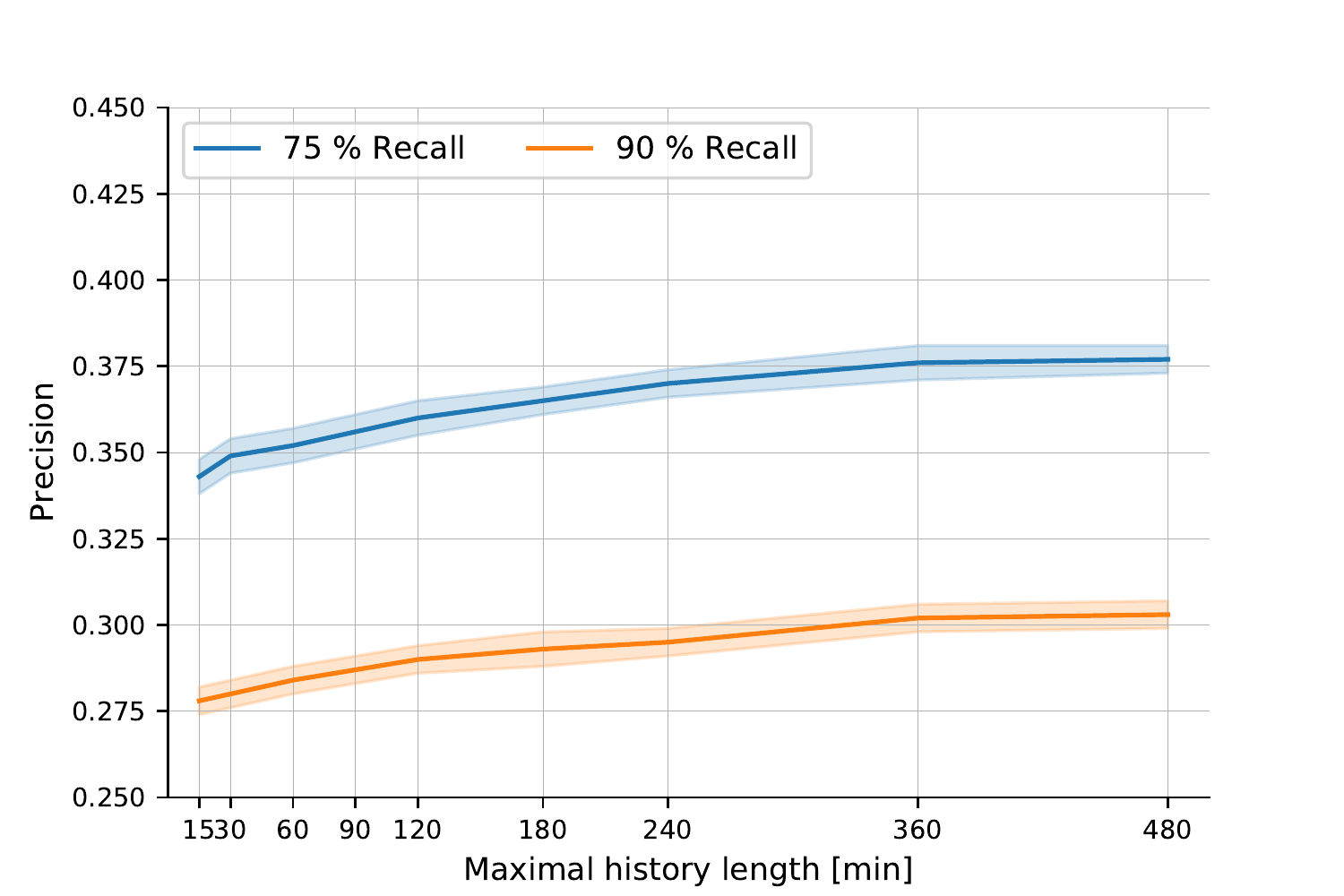}
\caption{Model performance conditional on maximum history length contained in the feature set.}
\label{fig:history-length}
\end{figure}

Given the benefits of complex waveform features, it is still unclear whether informative changes in pre-hypertensive patterns occur during 
the short- or also long-term history before the event. We tried to answer this question by ablating the set of multi-scale summary functions supported by our
framework. Our results (Fig. \ref{fig:history-length}) indicate that there seem to be no clear saturation effects when 
adding averages/trends/variability over additional length scales until a history length of 6 hours, when performance saturates.

\subsection*{Comparison of proposed model with baselines}

\begin{table}[h!]
\caption{Comparison of different machine learning methods applied to the optimal features found by SHAP analysis.}
\centering
\resizebox{0.49\textwidth}{!}{%
\begin{tabular}{llll}
\textbf{ML method} & \textbf{Prec@75Rec} & \textbf{Prec@90Rec} & \textbf{AUPRC} \\
\toprule
\textbf{LogReg} & \textbf{0.377} $\pm$ 0.001 & \textbf{0.303} $\pm$ 0.001 & \textbf{0.517} $\pm$ 0.003 \\ 
\textbf{MLP} & \textbf{0.373} $\pm$ 0.002 & \textbf{0.300} $\pm$ 0.002 & \textbf{0.514} $\pm$ 0.003 \\ 
\textbf{Tree} & 0.311 $\pm$ 0.001 & N/A & 0.375 $\pm$ 0.003 \\ 
\textbf{GradBoost} & 0.353 $\pm$ 0.002 & 0.276 $\pm$ 0.001 & 0.465 $\pm$ 0.003 
\end{tabular}}
\label{table:ml-model}
\end{table}

In a last step, we evaluated different machine learning methods fitted on the optimal features
and compared with two baselines from the literature in Tables~\ref{table:ml-model} and \ref{table:model-baselines}. 
We simulated the method of Hu et al. \cite{hu10_forecasting_icp_elevation} 
(\textbf{BL1: ICP morphology}) by computing medians of ICP pulse morphology metrics in the last 15 and 30 minutes, which is similar to their pre-hypertensive segment features. 
A second baseline implements the recently proposed method by 
Myers et al. \cite{myers16_predicting_intracranial_pressure} (\textbf{BL2: Last 2 points + Time to last crisis}) 
which uses as the three features the last 2 ICP values in a 30 minute window and the time since the last ICH event. If there was no such event, the last feature 
was set to a large symbolic value. Results show that the simplest machine learning model, i.e. \texttt{LogReg}, performed the best, 
while more complicated models like neural networks (\texttt{MLP}) or tree-based methods (\texttt{Tree}/\texttt{GradBoost}) provided no improvements. 
N/A is shown in Table~\ref{table:ml-model} if the recall could not be achieved by a machine learning method in all splits. Our proposed model 
significantly outperformed the two baselines, both for PR-based and ROC-based metrics that emphasive high sensitivities.

\begin{table}[h!]
\caption{Comparison of our proposed model with different baselines from the literature.}
\centering
\resizebox{0.49\textwidth}{!}{%
\begin{tabular}{llll}
\toprule
\textbf{Models} & \textbf{Prec@75Rec} & \textbf{Prec@90Rec} & \textbf{AUPRC} \\ 
\midrule
\textbf{Optimal} & \textbf{0.377} $\pm$ 0.001 & \textbf{0.303} $\pm$ 0.001 & \textbf{0.517} $\pm$ 0.003 \\ 
\textbf{BL1: Hu et al. \cite{hu10_forecasting_icp_elevation}} & 0.331 $\pm$ 0.001 & 0.274 $\pm$ 0.001 & 0.473 $\pm$ 0.003 \\ 
\textbf{BL2: Myers et al. \cite{myers16_predicting_intracranial_pressure}} & 0.338 $\pm$ 0.002 & 0.259 $\pm$ 0.001 & 0.484 $\pm$ 0.003 \\ 
\toprule
& \textbf{Spec@75Sens} & \textbf{Spec@90Sens} & \textbf{AUROC} \\
\toprule 
\textit{as above} & 0.653 $\pm$ 0.002 & 0.417 $\pm$ 0.003 & 0.771 $\pm$ 0.001 \\ 
 & 0.577 $\pm$ 0.002 & 0.336 $\pm$ 0.003 & 0.738 $\pm$ 0.001 \\ 
 & 0.602 $\pm$ 0.003 & 0.304 $\pm$ 0.003 & 0.746 $\pm$ 0.001 
\end{tabular}}
\label{table:model-baselines}
\end{table}

\subsection*{Ranking of most important physiological metrics}

By computing mean absolute SHAP values on the validation set in all 10 splits, we obtained a joint ranking of importance
of individual physiological metrics. The 20 most important features for predicting ICH are listed in Table \ref{table:most-important-features}. 
Features that have identical signatures but are computed over distinct scales, are not repeated. Instead, scales that belong 
to the top 100 features overall are listed in the last column. Features among the top 100 for special feature categories are listed in 
Tables \ref{table:feature-ranking-icp-abp-wave} (wICP/wABP waveform) and \ref{table:feature-ranking-index} (auto-regulation indices).

\begin{table}[h!]
\caption{Overview of 20 most important features for predicting ICH, identified by the SHAP analysis in the 10 splits}
\centering
\resizebox{0.49\textwidth}{!}{%
\begin{tabular}{llll}
\textbf{Rank} & \textbf{Feature descriptor} & \textbf{Important scales}\\
\midrule
1 & Med(IcpPulse\_Dias(\texttt{wICP})) & 480,360,30,15,60,180,240 \\ 
2 & Med(SpectralEnergy(\texttt{wICP})\_0-1Hz) & 360,480,30,240 \\ 
3 & Med(IcpPulse\_Mean(\texttt{wICP})) & 480,360,120,180 \\  
6 & Time since segment start & N/A \\ 
9 & Med(SpectralEnergy(\texttt{ECG})\_0-1Hz)  & 480,180,60 \\ 
15 & Med(AmpIndex(\texttt{ABPm},\texttt{CPP})) & 480 \\ 
17 & Med(SpectralEnergy(\texttt{wPLETH})\_2-3Hz) & 480 \\ 
18 & Med(SpectralEnergy(\texttt{wICP})\_9-12Hz) & 480 \\ 
19 & Med(SpectralEnergy(\texttt{wRESP})\_0-1Hz) & 480 \\ 
22 & Med(SpectralEnergy(\texttt{wPLETH})\_0-1Hz) & 480 \\ 
24 & Med(SpectralEnergy(\texttt{wRESP})\_1-2Hz) & 480 \\ 
26 & Med(IcpPulse\_Slope(\texttt{wICP})) & 480 \\ 
27 & Med(ShannonEntropy(\texttt{HR})) & 480 \\ 
28 & Iqr(SpectralEnergy(\texttt{ECG})\_0-1Hz)  & 360 \\ 
29 & Current \texttt{ICP} value & N/A \\ 
30 & Med(SpectralEnergy(\texttt{wRESP})\_2-3Hz) & 480 \\ 
31 & Iqr(SlowWaveIndex(\texttt{ICP})) & 360 \\ 
32 & Med(Med(\texttt{ICP})) & 480 \\ 
33 & Med(SpectralEnergy(\texttt{wICP})\_6-9Hz) & 480 \\ 
34 & Iqr(SpectralEnergy(\texttt{wPLETH})\_1-2Hz)  & 480 
\end{tabular}}
\label{table:most-important-features}
\end{table}

As a complementary analysis to feature ablation, we also analyzed which feature categories provided
important features according to rankings of mean absolute SHAP values. To enable an easier comparison, we computed the fraction of actual 
inclusions in the top 100 features (per split) over the 
number of theoretically possible inclusions. Results are summarized in Table \ref{table:feature-ranks-grouped}. Waveforms contribute more to
highly ranked features than time series, both in absolute and relative terms. In addition, several important features are
auto-regulation indices, spectral energies or morphological summaries. Finally, long-scale history summaries 
between 4 and 8 hours also provide many highly ranked features.

\begin{table}[h!]
\caption{Overall importance of feature categories evaluated by the number of inclusions in the top 100 
         features, across 10 splits}
\centering
\resizebox{0.49\textwidth}{!}{%
\begin{tabular}{lll}
\textbf{Feature descriptor} & Inclusion count & Normalized inclusion count \\
\midrule
\textbf{Physiological channel} & & \\
\texttt{wICP} & \textbf{369} & \textbf{0.050} \\ 
\texttt{ECG} & \textbf{344} & \textbf{0.051} \\  
\texttt{wABP} & \textbf{128} & 0.020 \\ 
\texttt{ICP} & 83 & 0.029 \\ 
\texttt{wRESP} & 79 & \textbf{0.047} \\  
\texttt{wPLETH} & 71 & 0.042 \\  
\texttt{CPP} & 60 & 0.023 \\  
\texttt{ABPm} & 53 & 0.022 \\  
\texttt{HR} & 28 & 0.029 \\ 
\texttt{ABPs} & 25 & 0.026 \\  
\texttt{ABPd} & 7 & 0.007 \\ 
\midrule
\textbf{Base feature function} & & \\
SpectralEnergy & \textbf{463} & \textbf{0.055} \\ 
IcpPulseMorph & \textbf{200} & 0.042 \\ 
Median & \textbf{65} & 0.034 \\  
Entropy & 56 & 0.029 \\  
AbpPulseMorph & 53 & 0.013 \\  
LineLength & 34 & 0.018 \\ 
TFIndex & 31 & 0.032 \\ 
SlowWaveIndex & 26 & \textbf{0.054} \\ 
AmpIndex & 21 & 0.022 \\ 
PrxIndex & 14 & \textbf{0.058} \\ 
Iqr & 13 & 0.007 \\ 
IaacIndex & 4 & 0.017 \\ 
PaxIndex & 4 & 0.017 \\  
RapIndex & 3 & 0.013 \\ 
\midrule
\textbf{Summary function} & & \\
Median & \textbf{715} & \textbf{0.076} \\ 
Iqr & 262 & 0.028 \\ 
Slope & 10 & 0.001 \\ 
\midrule
\textbf{History length [mins]} & & \\
480 Mins & \textbf{465} & \textbf{0.131} \\ 
360 Mins & \textbf{242} & \textbf{0.068} \\  
240 Mins & \textbf{104} & \textbf{0.029} \\  
180 Mins & 71 & 0.020 \\ 
120 Mins & 34 & 0.010 \\ 
60 Mins &  28 & 0.008 \\ 
15 Mins &  24 & 0.007 \\ 
30 Mins &  19 & 0.005 
\end{tabular}}
\label{table:feature-ranks-grouped}
\end{table}

\subsection*{Alarm timeliness before events}

\begin{figure}[h!]
\centering
\includegraphics[width=0.49\textwidth]{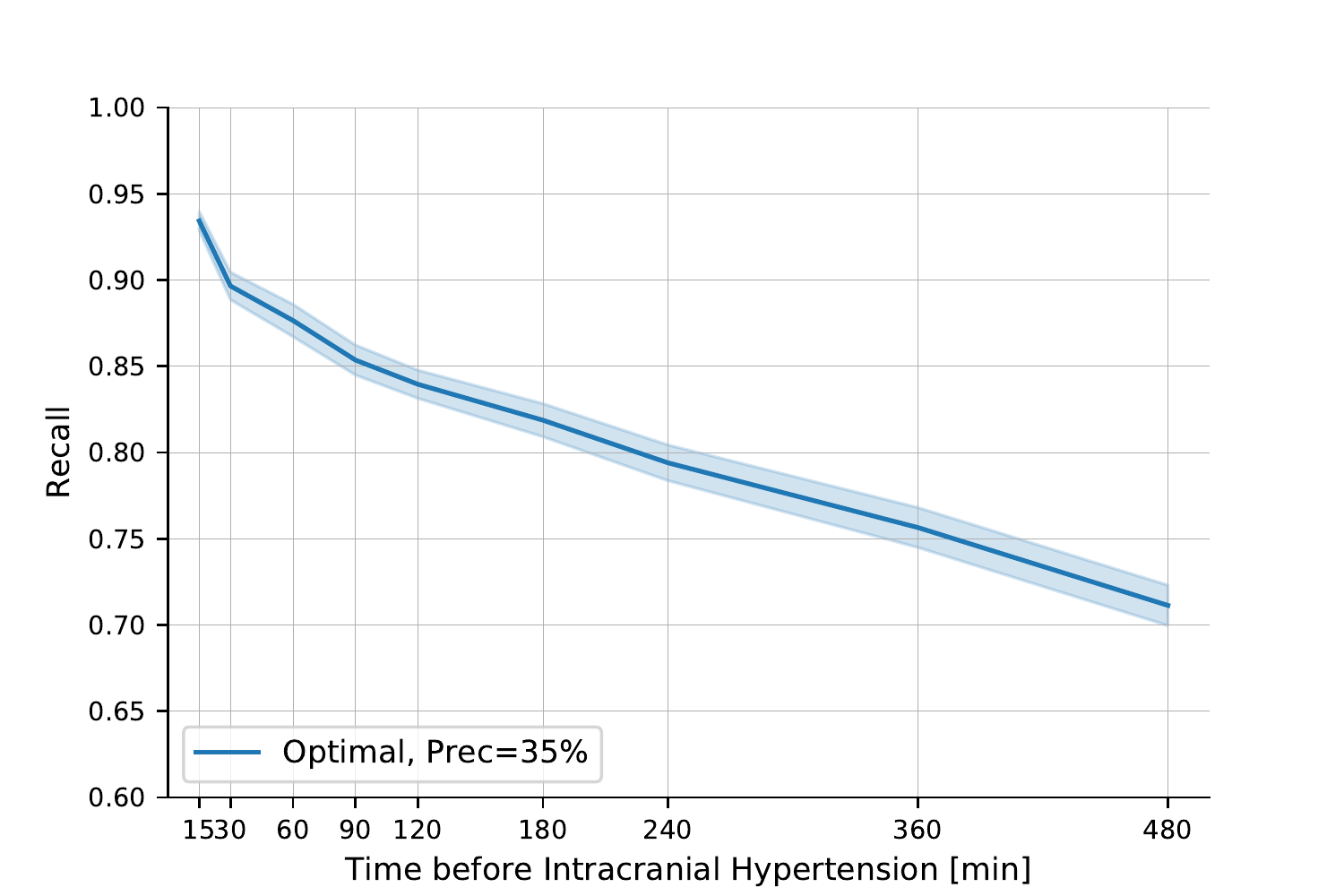}
\caption{Evaluation of alarm timeliness before a future ICH event, by time-before-event, of the proposed model.}
\label{fig:analysis-timeliness}
\end{figure}

The performance of the proposed model for a precision of 35\% is shown in Fig. \ref{fig:analysis-timeliness}. To provide more 
insights into the behavior of a derived alarm system in clinical settings, we have 
analyzed the recall of desired alarms before events, conditional on the time until the ICH phase starts. This 
measures the timeliness of alarms given a fixed model with a constant overall false alarm rate, which is a realistic
scenario of clinical implementation. We can observe a modest decay of alarm recall rates in 
the proposed model, which stay above 70\% even 8 hours prior to the event.

\begin{table}[h!]
\caption{Most important physiological metrics extracted from high-frequency ICP/ABP waveforms, among the top 100 features overall}
\centering
\resizebox{0.49\textwidth}{!}{%
\begin{tabular}{llll}
\textbf{Rank} & \textbf{Feature descriptor} & \textbf{Most important scales}\\ 
\midrule
1 & Med(IcpPulse\_Dias(\texttt{wICP})) & 480,360,30,15,60,180,240,120 \\ 
2 & Med(SpectralEnergy(\texttt{wICP})\_0-1Hz) & 360,480,30,240,120,180,60,15 \\ 
3 & Med(IcpPulse\_Mean(\texttt{wICP})) & 480,360,60,120,15,240 \\ 
18 & Med(SpectralEnergy(\texttt{wICP})\_9-12Hz) & 480,360 \\ 
26 & Med(IcpPulse\_Slope(\texttt{wICP})) &  480,180,360 \\ 
33 & Med(SpectralEnergy(\texttt{wICP})\_6-9Hz) & 480 \\ 
36 & Med(IcpPulse\_DecayTimeConst(\texttt{wICP})) & 480 \\ 
50 & Iqr(SpectralEnergy(\texttt{wICP})\_9-12Hz) & 480 \\ 
52 & Med(IcpPulse\_L2(\texttt{wICP})) & 480 \\ 
61 & Med(SpectralEnergy(\texttt{wICP})\_12-15Hz) & 480 \\ 
68 & Med(IcpPulse\_Curve1(\texttt{wICP})) & 480 \\ 
76 & Med(IcpPulse\_L3(\texttt{wICP})) & 480 \\ 
83 & Med(SpectralEnergy(\texttt{wICP})\_9-12Hz) & 360 \\ 
85 & Iqr(IcpPulse\_DP13(\texttt{wICP})) & 480 \\ 
93 & Iqr(SpectralEnergy(\texttt{wICP})\_12-15Hz) & 480 \\ 
99 & Med(IcpPulse\_DP3(\texttt{wICP})) & 480 \\ 
\midrule 
66 & Med(AbpPulse\_AverageLatency(\texttt{wABP})) & 480 \\ 
72 & Med(SpectralEnergy(\texttt{wABP})\_12-15Hz) & 480 \\ 
\end{tabular}}
\label{table:feature-ranking-icp-abp-wave}
\end{table}

\begin{table}[h!]
\caption{Most important cerebral auto-regulation indices extracted from time series and waveforms, among 
         the top 100 features overall}
\centering
\resizebox{0.49\textwidth}{!}{%
\begin{tabular}{llll}
\textbf{Rank} & \textbf{Feature descriptor} & \textbf{Most important scales}\\
\midrule
15 & Med(AmpIndex(\texttt{CPP},\texttt{ABPm})) & 480 \\ 
31 & Iqr(SlowWaveIndex(\texttt{ICP})) & 360,480 \\ 
39 & Med(PrxIndex(\texttt{ICP},\texttt{CPP},\texttt{ABPm})) & 480 \\ 
44 & Iqr(TFIndex(\texttt{wICP},\texttt{wABP})) & 480,360 \\ 
64 & Med(TFIndex(\texttt{CPP},\texttt{ABPm})) & 480,360 \\ 
77 & Iqr(SlowWaveIndex(\texttt{wICP})) & 480 \\ 
\end{tabular}}
\label{table:feature-ranking-index}
\end{table}
\section{Discussion}

We have designed and evaluated a prediction framework for acute intracranial hypertension events, which describes the 
neurological state using multi-scale descriptors of cerebral autoregulation indices, pulse morphology metrics,
spectral energies and statistical summaries. Alarms before critical events were retrieved up to 8
hours before the onset of ICH (Figure \ref{fig:analysis-timeliness}) with an overall recall 
of 90 \% at a precision of 30.3 \%. By mainly analyzing the system using recall/precision we have chosen metrics that more 
easily translate to the clinical deployment of alarm systems than ROC-based metrics. The achieved AUROC score of 0.771
for 8h forecasting is comparable to the work of Myers et al. \cite{myers16_predicting_intracranial_pressure}, which is to our knowledge the only published 
work with a similar forecasting horizon of 6 hours. Yet, direct comparison 
is not easily possible due to differences in cohort size (123 vs. 817 segments) and label definitions. In this work, we implemented 
the method in \cite{myers16_predicting_intracranial_pressure} to enable a controlled comparison on the 
same dataset, see Table \ref{table:model-baselines}.

Limitations of our study include the inability to match all 123 recording segments to the MIMIC-III clinical database, and hence take clinical context
information into account for label and experimental split definition. With respect to labels, by not taking clinical interventions into account, 
we suspect that we are predicting cases of acute intracranial hypertension that care providers did not anticipate in time before the patient entered
the ICH state, which corresponds to scenarios where an early warning system is used as a complementary decision support tool.
To analyze this effect in detail, prospectively collected data which documents all clinical interventions and
provides contextual information on the ICH events would be required. In our data splits, training, validation and test sets are guaranteed to be temporally disjoint, without interaction between labels and clinical features. Focusing on patients that have comprehensive monitoring with few
missing values in the data-set could bias the cohort towards more intensively monitored patients. We chose to apply
this criterion regardless because it allows more meaningful conclusions about the utility of different feature channels for predictive modeling.

A novel perspective on the design of ICH alarm systems is provided by the results in Table \ref{table:benefit-ts-wave}. While building a model just using 
averaged time series defining the event status (\texttt{ICP}) provides a good baseline performance, the 
inclusion of richer data modalities like waveforms substantially increased performance. Including 
high-frequency context information (\texttt{wABP}) in addition to \texttt{wICP} increased the performance further. This hints at new 
independent information in the ABP waveform and supports the importance of auto-regulation indices, which are functions of two waveforms simultaneously. 
However, data storage and computational cost associated with waveform data might be considerable.

To our knowledge, there is no previous work that assesses the relative merits of different data modalities for ICH prediction. It is an interesting finding that
only using waveforms performed better than only averaged time series, especially in light of recent related works, which found high performance using very simple models, e.g. using only minute-by-minute summaries of ICP.

Each individual feature category among auto-regulation indices, spectral energy, pulse morphology metrics provided marginal performance gains (Table \ref{table:base-summary-functions})
and is relevant for explaining predictions, as assessed using SHAP values (Table \ref{table:feature-ranks-grouped}).
This shows that complex pre-hypertensive patterns previously identified can translate into relevant machine learning features in our framework.

Our results (Figure \ref{fig:history-length}) show a clear trend between the length of considered history and prediction performance, which confirms the design principle of the 
multi-scale history, and also hints at the relative importance of long-scale physiological changes before hypertensive events, which could inform clinical studies. The same 
observation can be derived from the analysis of feature category importance in Table \ref{table:feature-ranks-grouped}, where a clear trend in importance from short-term 
to long-term features is visible.

The comparison of machine learning models (Table \ref{table:ml-model}) provides a pragmatic look at the relevance of the exact statistical learning method for predicting ICH. As has been observed also
for other prediction problems in health care, simple models perform surprisingly well, and are not outperformed by models with higher complexity. We suspect that, since the feature choices
already incorporate extensive domain knowledge, a simple model like logistic regression is powerful enough. Given the similar performance of \texttt{MLP} and \texttt{LogReg}, we did not
consider the construction of more complex architectures like RNNs or CNNs for this study, which are harder to interpret than classical models, where interpretation
of model features is an important focus of this study.
\section{Conclusion}

We presented an online machine learning and signal processing framework that forecasts onsets of acute intracranial 
hypertension up to 8 hours in advance. Using an extensive series of ablation studies we have shed light on the critical 
components of the framework. SHAP value analysis provided a second perspective on the importance
of different feature categories in explaining predictions as well as a ranking of discriminative pattern changes before
acute intracranial hypertension. Both perspectives highlight the importance of information derived from waveforms, 
which provided a substantial performance increase. Our method out-performed two baselines from the literature, which use ICP pulse morphology and 
3 simple features of the ICP time series, respectively.

Directions of future work includes more sophisticated artifact detection methods at the block level, to minimize the corruption of down-stream feature 
generation, which is highly sensitive to accurate input signals. Exploring the per-sample SHAP values could provide interpretable reasons for predictions of future ICH events, visualize
regions of interest in the history, flag abnormal physiological indices that could precede ICH, as well as generate hypotheses for future studies on the phenomenon. 
Furthermore, our method could be extended by providing a calibrated alarm system on top of the prediction scores, which triggers
alarms at the bedside as a function of sequences of prediction scores. This would be an important step towards clinical implementation of our proposed approach, which 
was conceived as a real-time algorithm that could directly use data streamed from sensors. Collecting prospective clinical data to refine
the labeling using information on clinical interventions as well as adding clinical co-variates like diagnosis or clinical note concepts could, we suspect, 
increase the performance of our model even further. Finally, we expect that our framework could be applied to predict other critical events 
occurring in the injured brain.

\appendices

\section*{Acknowledgments}

This work was supported by the Gebert-R\"uf Stiftung, Switzerland, under grant 
agreement GRS-025/14 and the Swiss National Science Foundation, under grant agreement 150640. 
MH was partially funded by the Grant No. 205321\_176005 “Novel Machine Learning Approaches for Data
from the Intensive Care Unit” of the Swiss National Science Foundation (to Gunnar R\"atsch). We acknowledge 
Emanuela Keller, head of the Neurocritical Care Unit at the University Hospital Z\"urich, Switzerland, for 
providing indispensible clinical insights and motivation for this work. MH gratefully acknowledges helpful
discussions with Panagiotis Farantatos, Viktor Gal, Stephanie Hyland, Xinrui Lyu, Ngoc M. Pham and Gunnar R\"atsch.

\bibliographystyle{IEEEtran}
\bibliography{IEEEabrv,../bibliography/references}

\end{document}